\documentclass[twocolumn,showpacs,amssymb,aps]{revtex4}
\usepackage{amsmath}
\usepackage{amsfonts}
\usepackage{amssymb}
\usepackage{graphicx}
\usepackage{epsfig}
\usepackage{dcolumn}
\usepackage{bm}

\def\lessim{\lower.5ex\hbox{$\; \buildrel < \over \sim \;$}}
\begin{document} \hbadness=10000
\topmargin -1.cm
\oddsidemargin = -0.4cm\evensidemargin = -0.1cm
\preprint{}

\title{Thermal reaction processes in a relativistic QED plasma drop}
\author{Inga Kuznetsova, Dieter Habs and Johann Rafelski}
\affiliation{Department of Physics, University of Arizona, Tucson, Arizona, 85721, USA}
\affiliation{Department f\"ur Physik der Ludwig-Maximilians-Universit\"at M\"unchen und
Maier-Leibnitz-Laboratorium, Am Coulombwall 1, 85748 Garching, Germany}

\date{October, 30, 2009}
\abstract {The equilibrium size and temperature limits of thermally and chemically 
equilibrated $e^+e^-\gamma$ plasma drops are investigated at a given energy content. 
For a plasma to be equilibrated it must be opaque to electron and photon interactions. 
The opaqueness condition is determined by comparing plasma size with the mean free 
electron and photon paths. We calculate those paths using thermal Lorentz-invariant 
reaction rates for pair production and electron (positron) and photon scattering. 
The range of the corresponding plasma temperature and size is evaluated numerically. 
Considering the energy and size we find that the opaque and equilibrated plasma drop 
may be experimentally attainable.

} }
\maketitle 

\section{Introduction}

 In~\cite{Kuznetsova:2008jt} 
we considered relativistic EP$^3$ ($e^-,e^+,\gamma$ plasma, 
{\bf e}lectron, {\bf p}ositron {\bf p}hoton {\bf p}lasma) in  equilibrium
 and have studied heavy (muon, pion) 
particle production rates. Here we extend considerably  these methods and study  
the production   and equilibration of the components $e^-,e^+,\gamma$ 
of the EP$^3$ plasma in order to explore the constraints on how such a  
plasma can be formed, and when it falls apart  into free-streaming constituents (freeze-out condition). We  evaluate 
constraints arising from the requirement that 
a plasma drop ought to be opaque to constituent particle scattering, 
and include in this consideration 
the constraint on size and temperature arising from laser pulse energy
available to form the plasma. Furthermore, we
can determine under what conditions the plasma drop will not only reach thermal 
(spectral) but also chemical (yield of particles) equilibrium.

In laboratory experiments involving, e.g. ultra intense 
laser pulses to form strong fields,  electron-positron pairs are produced; for review 
see~\cite{Mourou:2006zz,Marklund:2006my,Ruffini:2009hg}. They  
rescatter or annihilate and produce hard photons. In a large enough domain the $e^+e^-$
pairs and hard photons $\gamma$ will equilibrate thermally and chemically. This 
kinetic equilibration process  cannot
be resolved in the general study we present here. Instead,  we focus our attention on 
the properties of the $e^+e^-\gamma$  plasma drop near equilibrium 
in its final evolution stage. 
We are assuming EP$^3$  was formed in the vacuum state
(particle-antiparticle symmetry) and keeping the total zero spin.

In laboratory experiments an initially dense plasma fireball with radius $R_{\rm pl}$  will 
dilute, expanding at plasma speed of ``sound'', nearly speed of light. The lifespan  in a fast 
expanding plasma drop is 
$$\tau_{\rm pl}\simeq \frac{\sqrt{3}}{c}R_{\rm pl}.$$ 
We consider the plasma drop of finite size. However we neglect the energy loss 
from the radiation of photons from the plasma surface. We assume that if a plasma 
is opaque then the energy loss from a surface is insignificant during the plasma 
lifespan, which is defined by the laser pulse duration. If plasma is nonopaque, 
then energy loss occurs from the whole plasma volume, signaling the end of 
equilibrium and onset of transport regime not addressed here.

There are three different conditions which we can track in the plasma state
which are usually passed in sequence in time:
\begin{enumerate}
\item At highest density (i.e. temperature $T$) we 
reach particle yield  chemical equilibrium. 
This is the condition when the  $e^+e^-$-pair annihilation into 
and production by photons occur at equal rate.
\item At an intermediate density many scattering processes can occur, but not
as many as needed to maintain chemical  equilibrate. That is the opaqueness domain.

At this time also most photons will leave the plasma free-streaming  
\item At yet a lower  density, freeze-out of  $e^+, e^-$ occurs; that is, these
particles will leave the plasma without further rescattering.
\end{enumerate}
Let us look at how these conditions are obtained, in reverse order:
for the particle freeze-out (free-streaming) condition we consider 
\begin{equation}
L_i\simeq R_{pl},  \quad\mathrm{free-streaming}. \label{free-stream}
\end{equation}
Here we use the  thermal free path $L_{i}$, $i=e^-,e^+,\gamma$ which is computed
below,  and compare it to the plasma radius $R_{pl}$. 
For plasma to be opaque we require
\begin{equation}
L_i \leq \frac{R_{\rm pl}}{3}, \quad\mathrm{opaque}. \label{opcon}
\end{equation}
Here and below the factor 3 is a putative choice, made in view of other studies
of kinetic plasma dynamics.
The $e^+e^-$ pair and photon chemical 
equilibration condition is approximately the same as the
opaqueness condition, Eq.(\ref{opcon}) for photons 
participating in the pair production reaction Eq.(\ref{ggee1}):
\begin{equation}
L_{\gamma\gamma \rightarrow ee} \leq \frac{R_{\rm pl}}{3}, 
   \quad\mathrm{chemical\ equilibrium}. \label{cheqcon1}
\end{equation}
The free-streaming and opaqueness conditions are up  to numerical evaluation the same.

\begin{figure}[t]
\centering
\includegraphics[width=9.3cm]{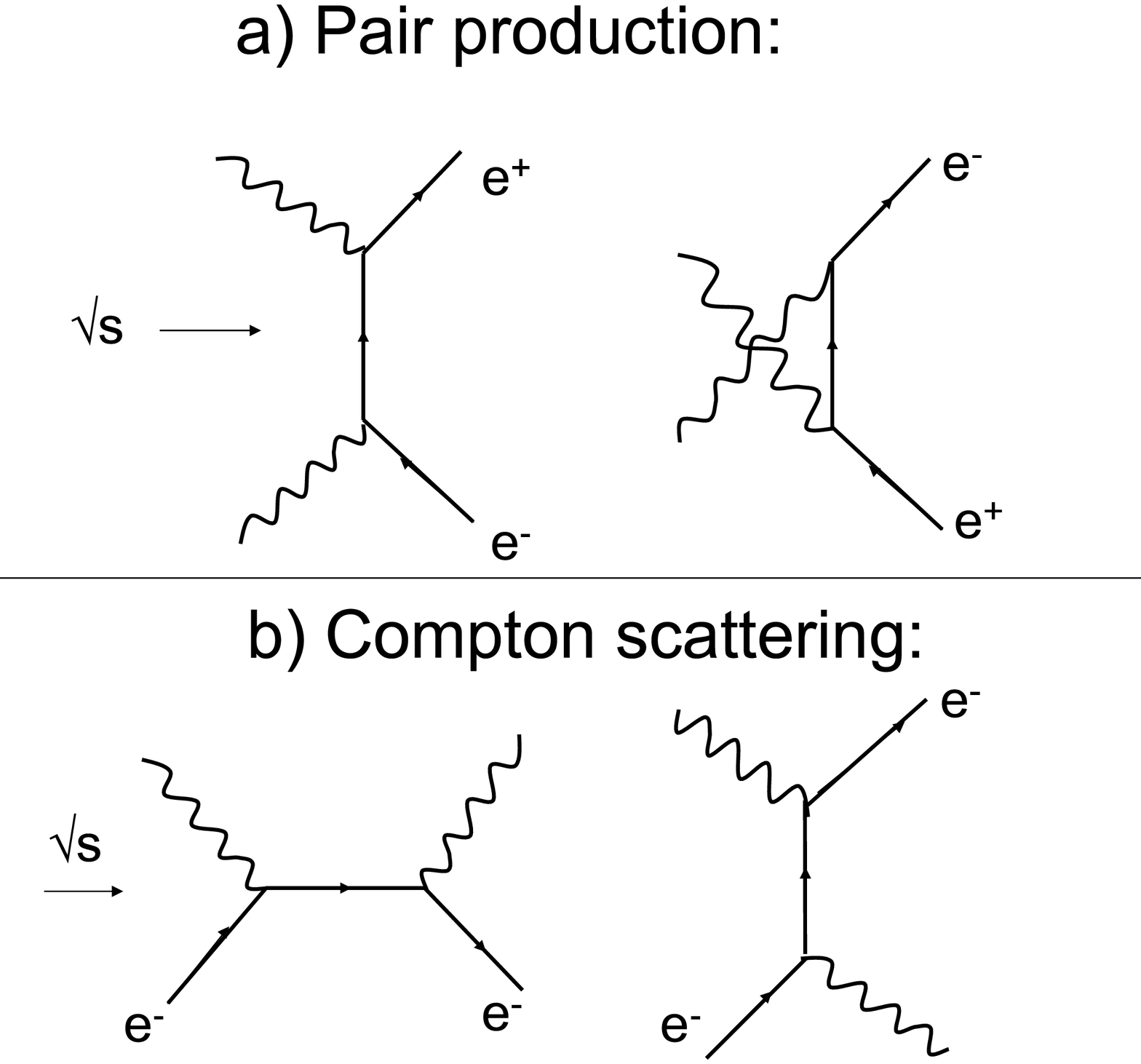}
\caption{\small{Feynman diagrams for 
     (a) the creation of the electron-positron pair on top  
and  (b) Compton scattering on the bottom,   
where incoming particles are on the left side.
}} \label{eeprod}
\end{figure}

We consider the mean free path $L_{\gamma}$ and $\gamma$ equilibration  
by Dirac pair production/annihilation reaction:
\begin{equation}
\gamma + \gamma \leftrightarrow e^+ + e^-, \label{ggee1}
\end{equation}
and  by Compton scattering:
\begin{equation}
\gamma+e^{\pm} \leftrightarrow \gamma + e^{\pm}, \label{ge}
\end{equation}
for which the Feynman diagrams are shown in Fig. \ref{eeprod}.

\begin{figure}[t]
\centering
\includegraphics[width=9.3cm]{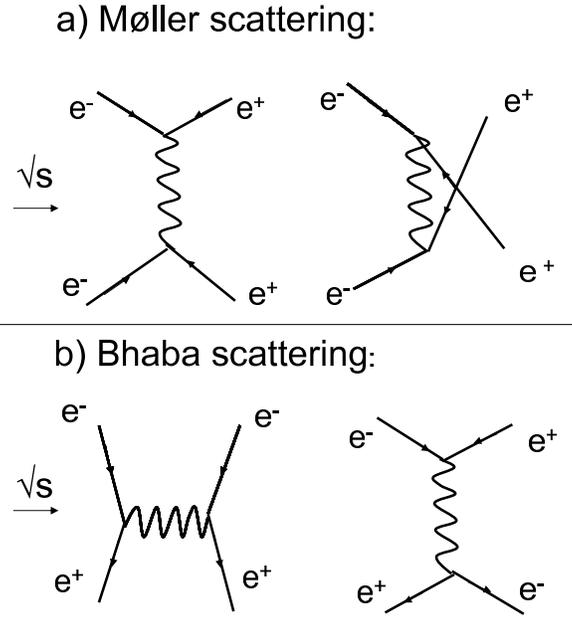}
\caption{\small{Feynman diagrams for 
    (a) M{\o}ller scattering on top   
and (b) Bhabha scattering on bottom,   
where incoming particles are on the left side.
}} \label{eediagramsp}
\end{figure}

However, the evaluation of chemical equilibration
of electrons and  positrons is in essence different, these particle  
will continue to rescatter  even 
after all photons have left, and  $e^-,e^+$
can pair annihilate into photons. 
We  obtain the mean free path $L_{e^-,e^+}$ and the $e^+e^-$ pair equilibration   
considering M{\o}ller:
\begin{equation}
e^{\pm}e^{\pm} \leftrightarrow e^{\pm}e^{\pm},\label{Moller}\\
\end{equation}
and Bhabha scattering:
\begin{equation}
e^{+}e^{-} \leftrightarrow e^{+}e^{-},\label{bhaba}
\end{equation}
for which the Feynman diagrams are shown in Fig. \ref{eediagramsp}.

We evaluate using a Lorentz covariant 
procedure the invariant reaction rates using methods
developed first for strange quark pair production 
in the quark gluon plasma~\cite{Koch:1986ud, Matsui:1985eu}. 
We extend these methods to scattering reactions  
which do not change the number of particles.

Our kinetic equations describing the rates and particle multiplicity evolution 
are obtained from the Boltzmann equation, integrated over the momentum part of the phase 
space, assuming thermal distribution. As a result, we obtain thermally 
averaged reaction relaxation time. This approach  is appropriate  to 
identify the domain of $T$ in which reactions are fast  enough to maintain 
thermal and chemical equilibrium. However, a future  study will need to address  
how the thermal distribution is established. We also see a need to consider in
a more complete transport approach the motion  of  particles found in high energy
tail of distribution, which will be first to escape the plasma drop. However, 
these particles contribute little to the bulk properties.

We observe that  the rates for reactions considered 
here had been  obtained by the other methods, using 
cross sections~\cite{WeaverT:1976} and applied 
for astrophysical environments~\cite{Svensson:1982hz}. 
Our results differ from those presented in~\cite{Svensson:1982hz}, 
which found that under certain conditions plasma equilibration 
occurs primarily by pair production, and the Compton scattering has 
small effect. We show that the Compton scattering is always the 
dominant process near equilibrium. The work to reconcile~\cite{Svensson:1982hz}   
with our covariant results continues. 
One possible source of discrepancy is that in~\cite{WeaverT:1976} 
the Lorentz-invariant form of distribution function
[see Eq.(\ref{fstat})] is not used in the rate evaluation procedure  which involves 
a later  change of reference frames. Therefore our Lorentz covariant results  are
different and amend this earlier work.

We do not consider the three-body reactions. Their rates are 
proportional to an additional factor $\alpha/\pi$, and thus in 
general are slower  than two-body reactions considered here.
We find that Bhabha and M{\o}ller  processes are much faster than 
other reactions considered. However, if Bremsstrahlung is 
about $\alpha  = 1/137$ times slower than Bhabha and M{\o}ller 
scatterings, it could still be faster, or of the same order 
of magnitude, as Compton and pair production. Bremsstrahlung 
and the related backward processes could therefore somewhat 
shorten photon equilibration, perhaps visible in some domain 
of $T$. We leave this to future study. 

Our paper is organized as follows: In section~\ref{plpr1} we review the physical  
properties (subsection \ref{plpr1A}) of EP$^3$ plasma, and present (subsection~\ref{plpr1B})
the master equations for chemical equilibration of the plasma components under the assumption 
that particles are in thermal equilibrium. We introduce  $\tau^{ch}_{\gamma\gamma}$,
the relaxation time for photon chemical equilibration, in terms of thermal reaction rates,
and relate $c\tau^{\rm ch}_{\gamma\gamma}$  to the mean free path for pair production. 

In Sec.~\ref{plpr2} we  obtain the analytical expressions for the thermal
mean free paths for scattering in EP$^3$ plasma. 
In Sec. \ref{plpr2A} we review equations for mean free paths 
and  Lorentz-invariant rate calculations.  
In Sec. \ref{plpr2B} we review cross sections 
for pair production and annihilation and Compton scattering 
as functions of the energy of interacting particles.
In Sec. \ref{plpr2C} we obtain the Lorentz-invariant 
rate for pair production  and annihilation.
Similar invariant rate calculations were done before for the strange quark pair 
production in Refs.~\cite{Koch:1986ud} and \cite{Matsui:1985eu}. We perform 
the required integration by a method similar to ~\cite{Matsui:1985eu}, and muon pair
production~\cite{Kuznetsova:2008jt} we considered earlier. In the following sections 
we extend that method to Compton (Sec. \ref{plpr2D}),    
M{\o}ller  and Bhabha (Sec. \ref{plpr2E}) invariant rates. To remove the 
Coulomb divergence in integrals we introduce a thermal screening photon (Debye) 
mass, induced by plasma oscillations. 

Sec.~\ref{numres} contains  our numerical results. In Sec.~\ref{Thompson} we 
look in depth at photon-electron interactions and the Thompson limit. In Sec.~\ref{Paths}
we evaluate the   photon and electron scattering lengths, and discuss maximum possible plasma size at 
a given energy. This evaluation is carried out taking account of all  quantum physics 
Fermi and Bose statistic effects
in order to be applicable in the high density domain,
and we apply the Lorentz-invariant rates method which include medium  quantum effects. 
We connect our results to the Boltzmann limit in Sec. \ref{oldappendix}.

In Sec.~\ref{conc} we discuss the  physical consequences,
applications, and present our conclusions. 

\section{Statistical properties of EP$^3$ plasma}\label{plpr1}
\subsection{Relativistic Gas}\label{plpr1A}
Up to   QED interaction ${\cal O}(\alpha)$ effects we can describe
the particle content in a plasma using, respectively, relativistic 
covariant Fermi and Bose momentum distributions:
\begin{equation} \label{fstat}
f_{e^\pm}  = \frac{1}{\Upsilon_e^{-1}e^{(u\cdot p_{e}\pm\nu_e)/T} +1},\quad
f_{\gamma}  =  \frac{1}{\Upsilon_{\gamma}^{-1}e^{u\cdot p_{\gamma}/T} -1},
\end{equation}
$\Upsilon_{i}$ ($i=e$, $\gamma$) is the fugacity of a given particle. 
When $\Upsilon_{i} = 1$, the particle yield is in chemical equilibrium.
For $\Upsilon_{i}=1$ this distribution maximizes the entropy content 
at a fixed particle energy~\cite{Letessier:1993qa}.

The Lorentz-invariant exponents involve the scalar product of the 
particle 4-momentum $p^{\mu}_{i}$ with the
local 4-vector of velocity $u^{\mu}$. In the absence of matter 
flow we have in the rest frame of the thermal bath (laboratory frame)
\begin{equation}
u^{\mu}=\left(1,\vec{0}\right),\label{4v}  \qquad p^{\mu}_{i}=\left(E_i, \vec{p_i}\right).
\end{equation}
When the electron chemical potential $\nu_e$ is small, $\nu_e\ll T$,
the number of particles and antiparticles is the same,
$n_{e^-}=n_{e^+}$. Physically, it means that the number of $e^+e^-$
pairs produced dominates the residual matter electron yield. Here we
will set $\nu_e=0$ , and will consider elsewhere the case for very
low density plasma where chemical potential may become important.

\begin{figure}[!hbt]
\centering
\includegraphics[width=8.6cm]{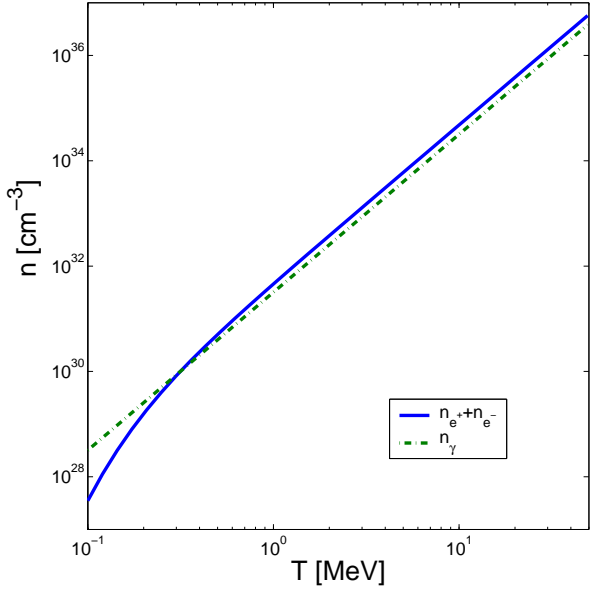}
\caption{\small{Electron-positron (blue solid line) and 
photon (green dash-dotted line) densities as functions of temperature.}} \label{neg}
\end{figure}

The particle density and plasma energy density can be
evaluated using relativistic expressions:
\begin{equation}
n_i = \int g_i f_i(p)d^3p;\,\,\,\,\,E = \int \sum_i g_iE_if_i(p)d^3pV,  \label{Etherm}
\end{equation}
where $E_i=\sqrt{m_i^2+\vec p^{\,2}}$, $f_i(p)$ is the  momentum
distribution of the particle $i \in \gamma$, $e^{\pm}$,
and $g_i$ is its degeneracy; $g_i=2$ for photons, electrons and positrons.
In Fig.~\ref{neg} we present electron-positron and photon 
densities as functions of temperature. At our range of 
temperatures their values  are much larger than atomic 
density, $10^{23}$ cm$^{-3}$.

After integration we obtain
\begin{equation}\label{SB}
\frac{\cal E}{V}= \epsilon   = \frac{\pi^2}{30}g(T)T^4.
\end{equation}
At a temperature $T\ll m_e$ we only have truly massless photons and $g(T) = 2_\gamma$.
Once the temperature approaches and increases beyond $m_e$,
we find  $g\simeq g^\prime(T) \simeq 2_\gamma+(7/8)(2_{e^-}+2_{e^+})=5.5$.

In a classical case, $\Upsilon << 1$, there is no difference between Bose and Fermi
particles. We have for massless particles
($m/T \rightarrow 0$)
\begin{equation}
E = 3NT, \qquad N=\Upsilon\frac{g}{\pi^2}T^3V, \label{energ}
\end{equation}
where $g = 6$. 

\subsection{Approach to equilibrium}\label{plpr1B}

In our approach new photons are produced by a pair annihilation 
reaction, Eq (\ref{ggee1}). If we assume that thermal equilibrium 
is established faster than chemical, then
the photon density evolution equations and chemical equilibrium conditions are
similar to those for muon production, studied in~\cite{Kuznetsova:2008jt}:
\begin{equation}
\frac{1}{V}\frac{dN_{\gamma}}{dt} = (\Upsilon_{e}^2
-\Upsilon_{\gamma}^2){R_{\gamma\gamma \leftrightarrow e^+e^-}} ,
\label{gammapr}
\end{equation}
where 
\begin{equation}
R_{\gamma\gamma\leftrightarrow e^+e^-} = 
  \frac{1}{\Upsilon^2_{\gamma}} \frac{dW_{\gamma\gamma 
       \rightarrow e^+e^-}}{dVdt}= \frac{1}{\Upsilon^2_{e}} \frac{dW_{e^+e^- 
           \rightarrow \gamma\gamma}}{dVdt}, \label{detbaleq}
\end{equation}
${dW_{\gamma\gamma \rightarrow e^+e^-}}/{dVdt}$ and ${dW_{e^+e^- 
          \rightarrow \gamma\gamma}}/{dVdt}$ 
are Lorentz-invariant rates for pair production and annihilation 
reactions, respectively. Rate $R_{\gamma\gamma\leftrightarrow e^+e^-}$ does not 
depend on $\Upsilon_i$ in limit of a classical Boltzmann distribution.
The EP$^3$ plasma is in {\em \underline{relative} chemical equilibrium} for
$\Upsilon_{\gamma}<1$ when
\begin{equation}
\Upsilon_{e} = \Upsilon_{\gamma}. \label{eppeq}
\end{equation}
In relative equilibrium at a given instant the considered reaction, 
for example (\ref{ggee1}), is in equilibrium.
However for other reactions this condition may not be an 
equilibrium  (for example for reaction $ee\leftrightarrow ee\gamma$) 
and then the system can move to another relative equilibrium condition 
with time. The relative chemical equilibrium occurs when crossing 
different time scale governs relevant ``chemical'' reactions. 
At $\Upsilon_{e, \gamma} \to 1$ (and for all others particles 
in the system, if any, $\Upsilon_i \to 1$) all reactions are 
in equilibrium and the system achieves full chemical equilibrium.

\begin{figure}
\centering
\includegraphics[width=8.6cm,height=8.5cm, height=8.5cm]{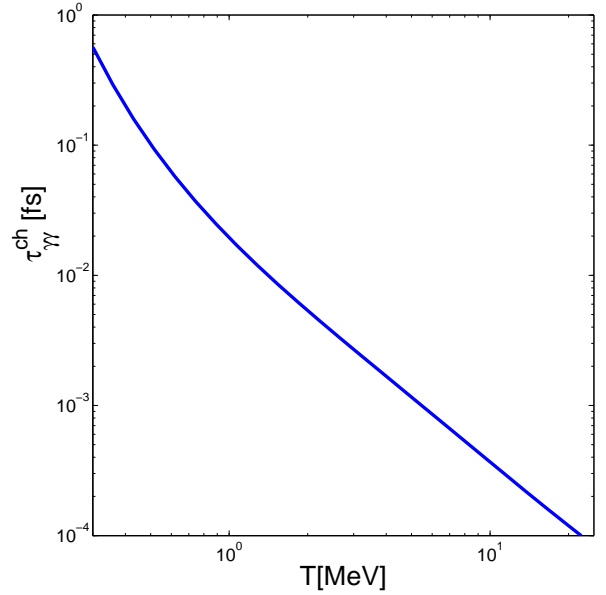}
\caption{\small{The relaxation time for pair and 
photon chemical equilibration.}} \label{taugg2}
\end{figure}

We introduce pair production relaxation time defined by:
\begin{equation}
\label{taugg} \tau^{ch}_{\gamma\gamma} =
\frac{1}{2\Upsilon_e}\frac{dn_{\gamma}/d\Upsilon_{\gamma}}{R_{\gamma\gamma
\leftrightarrow e^+e^-}}, 
\end{equation}
then for the simplest case $T(t)=$ Const. and
$R(t)=$ Const. the equation for $\Upsilon_{\gamma}$ is
\begin{equation}
\frac{d\Upsilon_{\gamma}}{dt} = (\Upsilon_e^2
-\Upsilon_{\gamma}^2)\frac{1}{2\Upsilon_e\tau^{ch}_{\gamma\gamma}}. \label{upgeq}
\end{equation}

The relaxation time vs temperature for photon and $e^+e^-$ 
pair chemical equilibration is shown in figure~\ref{taugg2}.
The rates and relaxation times are discussed in depth 
in~\cite{Kuznetsova:2008jt}. 

To understand better the choice of definition of relaxation 
time $\tau^{ch}_{\gamma\gamma}$, Eq.(\ref{taugg}), we assume 
$\Upsilon_e =const.$ and  introduce variable 
$\gamma = \Upsilon_{\gamma}/\Upsilon_{e}$, which 
shows deviation from chemical equilibrium. Then the equation for $\gamma$ is
\begin{equation}
\frac{d\gamma}{dt} = (1 - \gamma^2) \frac{1}{2\tau^{ch}_{\gamma\gamma}}.
\label{muev2}
\end{equation}
The relaxation time is defined as the time for particle 
multiplicity to reach the magnitude of equilibrium value.
 Note that the relaxation time for a particle (in this 
example photon) production in a two-to-two-particles 
reaction, Eq.(\ref{taugg}), changes by a factor
$\Upsilon_i^{-1}$, where $i$ denotes the initial particle in a reaction
(here $e^{\pm}$). The physical reason why the relaxation time is proportional to
$\Upsilon_i^{-1}$  is that the collision rate drops by this
 factor, due to reduced density of plasma.

In the simple case considered here, $e^+e^-$ pair production 
and annihilation are chemically equilibrated when condition 
Eq.(\ref{cheqcon1}) is satisfied.

\section{Mean free paths of photon and electron (positron) in $e^+e-\gamma$ plasma}\label{plpr2}

\subsection{Reaction rate and mean free paths}\label{plpr2A}

In order to be in thermal and chemical equilibrium plasma must be opaque for
the reactions, which establish this equilibrium. The major reactions which
may establish thermal and/or chemical equilibrium between photons and $e^+e^-$ pairs
are Compton scattering and pair production and annihilation.  $e^+e^-$ can also participate
in M{\o}ller scattering Eq.(\ref{Moller}) and Bhabha scattering Eq.(\ref{bhaba}).

Here we consider the reaction
\begin{equation}
1+2 \leftrightarrow 3+4, \label{1234r}
\end{equation}

For reaction Eq.(\ref{1234r}), the thermally averaged cross section is
\begin{equation}
\left\langle v\sigma_{12\rightarrow 34}\right\rangle  
= \frac{\Upsilon_1\Upsilon_2R_{12\leftrightarrow 34}}{n_1n_2}, \label{thcr}
\end{equation}
(velocity $v=c$ for photons scattering, we take $c=1$) and 
the mean free path is
\begin{equation}
L_{1(2)} = \frac{1}{n_{2(1)} \left\langle v\sigma_{12\rightarrow 34}\right\rangle}
=\frac{n_{1(2)}}{\Upsilon_{1}\Upsilon_{2}R_{12\leftrightarrow 34}}, \label{lgg}
\end{equation}
where $v$ is the relative velocity of interacting particles and $\sigma_{12\rightarrow 34}$ is
cross section.  

Rate $R$ is connected to Lorentz-invariant reaction rates 
by Eq.(\ref{detbaleq}). The equation for rate $R$ is \cite{Koch:1986ud, Matsui:1985eu}
\begin{eqnarray} 
&&R_{12 \leftrightarrow 34}=
\frac{1}{1+I}\frac{g_{\gamma}^2}{(2\pi)^8}\int\frac{d^{3}{p_{1}}}{2E_{1}} \int\frac{d^{3}{p_{2}}}{2E_{2}}
\int\frac{d^{3}{p_{3}}}{2E_{3}}
\int\frac{d^{3}{p_{4}}}{2E_{4}} \times
  \nonumber\\
&&\times \delta^{4}\left(p_{1} + p_{2} -
p_{3} - p_{4} \right) \sum_{\rm spin}\left|\langle
p_{1}p_{2}\left| M_{12 \rightarrow
34}\right|p_{3}p_{4}\rangle\right|^{2}\times \nonumber\\
&& f_{1}(p_{1})f_2(p_{2})f_{3}(p_{3})f_{4}(p_{4})
(\Upsilon_{1}\Upsilon_2\Upsilon_{3}\Upsilon_{4})^{-1}e^{\frac{u\cdot(p_{1} +
p_{2})}{T}},\label{m2ggee1}
\end{eqnarray}
where $I=1$, if there are identical particles in the initial or final state. 

Rate $R_{12\leftrightarrow 34}$ is the same for reactions in both directions,
because of time reversal symmetry of the matrix element:
\begin{equation}\left|\langle
p_{1}p_{2}\left| M_{12 \rightarrow
34}\right|p_{3}p_{4}\rangle\right|^{2} = \left|\langle
p_{3}p_{4}\left| M_{34 \rightarrow
12}\right|p_{1}p_{2}\rangle\right|^{2}.
\end{equation}
In the evaluation of matrix element we will use Mandelstam 
variables: s, u, and t.  For reaction (\ref{1234r}) we have
\begin{equation}
s = (p_1+p_2)^2; \quad u = (p_3-p_2)^2; \quad t=(p_3-p_1)^2; \label{mv}
\end{equation}
and $s+u+t=m_1^2+m_2^2+m_3^2+m_4^2$. 
  
If we exchange particles 2 and 3 as $1+3 \leftrightarrow 2+4$ 
or $p_2 \leftrightarrow p_3$, the matrix element for this reaction 
can be obtained from the matrix element for reaction 
Eq.(\ref{1234r}) by crossing s and t:
\begin{equation}
\left| M_{12 \rightarrow 34}(s,t,u)\right|^{2} 
= \pm \left|M_{13 \rightarrow 24}(t,s,u)\right|^{2}.\label{stcr}
\end{equation}
The sign before the matrix element may change to opposite depending 
if one of the particles, 3 or 2, changes from particle to antiparticle
after crossing.

For example, for pair production and annihilation and Compton 
scattering particles 1,2 are photons and particles 3,4 are electron and positron.
The matrix elements for these reactions is defined by Feynman 
diagrams in Figs.~\ref{eeprod}(a) and (b), respectively. 
The matrix element averaged over final states spins for pair 
production and annihilation is~\cite{Aksenov:2009dy}, \cite{Halzen:1984mc}
\begin{widetext}
\begin{equation}
|M_{\gamma \gamma \leftrightarrow  ee}|^2
 =  32 \pi^2 \alpha^2 \left(-4\left(\frac{m^2}{m^2-t}
 +\frac{m^2}{m^2-u}\right)^2+ 
 \frac{4m^2}{m^2-t}+\frac{4m^2}{m^2-u} +
 \frac{m^2-u}{m^2-t} +\frac{m^2-t}{m^2-u}\right), \label{m2ggee}
\end{equation}

The matrix element for Compton scattering is
\begin{equation}
|M_{\gamma e \leftrightarrow \gamma e}|^2
 =  32 \pi^2\alpha^2\left( 4\left(\frac{m^2}{m^2-s}
 +\frac{m^2}{m^2-u}\right)^2- 
 \frac{4m^2}{m^2-s}-\frac{4m^2}{m^2-u} -
 \frac{m^2-u}{m^2-s} -\frac{m^2-s}{m^2-u}\right), \label{m2comp}
\end{equation}
The matrix elements for these reactions are
connected by Eq.(\ref{stcr}).

Similarly, crossing s and u allows us to find
\begin{equation}
\left| M_{12 \rightarrow 34}(s,t,u)\right|^{2} = \pm \left| M_{23 
\rightarrow 14}(u,t,s)\right|^{2}.\label{sucr}
\end{equation}
The matrix elements for M{\o}ller and Bhabha scattering
 (diagrams are in Figs.~\ref{eediagramsp} (a) and (b), 
respectively) are connected by Eq.(\ref{sucr}). 
The M{\o}ller scattering matrix element is 
(similar to~\cite{Biro:1990vj}) 
\begin{equation}
|M_{e^{\pm}e^{\pm}}|^{2}=2^{6}\pi^{2}\alpha^{2}\left\{
\frac{s^{2}+u^{2}+8m^{2}(t-m^{2})}{2(t-m^2_{\gamma})^{2}}  +
\frac{{s^{2}+t^{2}}+8m^{2}%
(u-m^{2})}{2(u-m_{\gamma}^2)^{2}} + \frac{\left( {s}-2m^{2}\right)\left({s}-6m^{2}\right)}
  {(t-m_{\gamma}^2)(u-m_{\gamma}^2)} \right\}, \label{M_fi_m}%
\end{equation}
To obtain the Bhabha scattering matrix element we need 
to cross u and s, Eq.(\ref{sucr}) in M{\o}ller scattering matrix element (\ref{M_fi_m}).
We obtain
\begin{equation}
 |M_{e^+e^-}|^{2}=2^{6}\pi^{2}\alpha^{2}
\left\{
\frac{s^{2}+u^{2}+8m^{2}(t-m^{2})}{2(t-m^2_{\gamma})^{2}}  +\frac{u^{2}+t^{2}+8m^{2}%
(s-m^{2})}{2(s-m^2_{\gamma})^{2}}  +   \frac{\left({u}-2m^{2}\right)\left({u}-6m^{2}\right)}
   {(t-m^2_{\gamma})(s-m^2_{\gamma})} \right\}  , \label{M_fi_b}%
\end{equation}
and the photon thermal (Debye) mass $m_{\gamma}$ will be defined in Sec.~\ref{plpr2E}.

\subsection{Pair annihilation and production and Compton scattering cross sections}\label{plpr2B}

In this section we consider cross sections for pair annihilation 
and production and Compton scattering
as functions of energy.
The cross section $\sigma(s)$ can be obtained by averaging the matrix 
element over the $t$ variable. Similar calculations were done 
in~\cite{Combridge:1978kx} for heavy quarks production 
in quark gluon plasma. In the case of Compton scattering we obtain
\begin{eqnarray}
\sigma_{\gamma e \leftrightarrow \gamma e}(s)&=& 
  \frac{1}{16\pi(s-m^2)^2}\int_{-(s-m^2)^2/s}^0 dt|M_{\gamma e^{\pm}}|^{2} \nonumber\\[0.4cm]
 &=&\frac{2\pi\alpha^2}{(s-m^2)^2}\left(\frac{2(s-m^2)^2s-(s-m^2)^3}{2s^2}+8m^2-
 \left(\frac{8m^4}{s-m^2}+4m^2-(s-m^2)\right)\ln{\frac{s}{m^2}}
 \right)\label{crcom}
\end{eqnarray}
For pair production and annihilation we have (similar results as in~~\cite{Jauch:1976})
\begin{eqnarray}
\sigma_{\gamma\gamma \rightarrow ee}(s) &=& 
\frac{1}{16\pi s^2}\int_{m^2-s(1+\sqrt{1-4m^2/s})/2}^{m^2-s(1-\sqrt{1-4m^2/s})/2} dt
    |M_{\gamma\gamma \leftrightarrow ee}|^{2}\nonumber\\[0.4cm]
&=&\frac{4\pi\alpha^2}{s^3}\left(\left(s^2+4m^2s-8m^4\right)
   \ln{\frac{\sqrt{s}+\sqrt{s-4m^2}}{\sqrt{s}-\sqrt{s-4m^2}}}
      -\left(s+4m^2\right)\sqrt{s(s-4m^2)}\right).\label{crggee}\\
\nonumber\\[0.4cm]
\sigma_{ee \rightarrow \gamma\gamma}(s) &=& \frac{1}{32\pi s(s-m^2)}\int_{m^2-s(1+\sqrt{1-4m^2/s})/2}^{m^2-s(1-\sqrt{1-4m^2/s})/2} dt
  |M_{\gamma\gamma \leftrightarrow ee}|^{2}\nonumber\\[0.4cm]
&=&\frac{2\pi\alpha^2}{s^2(s-m^2)}\left(\left(s^2+4m^2s-8m^4\right)
   \ln{\frac{\sqrt{s}+\sqrt{s-4m^2}}{\sqrt{s}-\sqrt{s-4m^2}}} 
               -\left(s+4m^2\right)\sqrt{s(s-4m^2)}\right).
\label{creegg}
\end{eqnarray}
\end{widetext} 

Note that there is an extra 1/2
factor in the equation for pair annihilation because we have two
identical particles; that is a symmetrical wave function in the final state. Similarly,
in the reaction rate $R$ we add an additional factor 1/2 when there are initial
identical particles. In the backward reaction there is also a factor 1/2
from the definition of cross section. Therefore the rate is symmetrical in both reaction
directions: pair formation and annihilation.

\begin{figure}
\centering
\includegraphics[width=8cm,height=8cm]{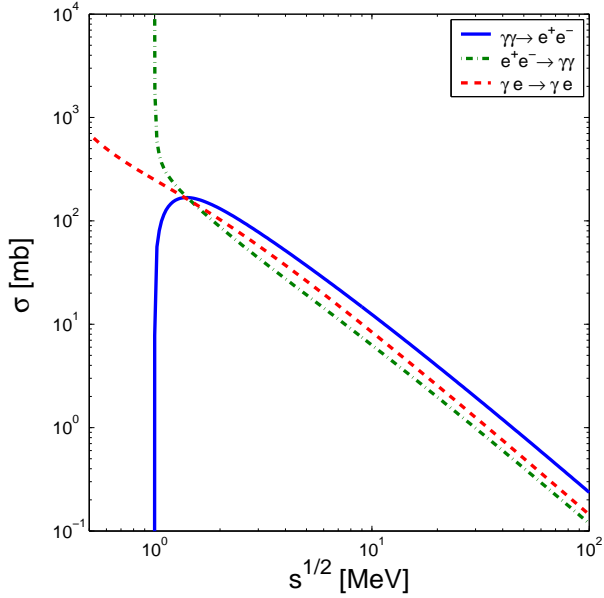}
\caption{\small{The cross sections for pair production 
(solid blue line) and annihilation (dash-dotted green line) 
in the center of mass frame, and for Compton scattering 
(dashed red line) in the electron rest frame are shown \
as functions of total energy of interacting 
particles $E_{\rm tot}$.}} \label{sigmaeg}
\end{figure}

In Fig.~\ref{sigmaeg} we show cross sections
(\ref{crcom})-(\ref{crggee}) as functions of total energy of
particles in reaction $s^{1/2}$ (center of mass frame) for pair
production (solid blue line), annihilation (dash-dotted green line) and  for Compton
scattering (dashed red line). 

At high energy $E_{\rm tot}>>m$ the cross section dependence on energy is a power law, 
\begin{equation}
\sigma = A\left(\frac{E_{\rm tot}}{1\,{\rm MeV}}\right)^{N} \label{pl}.
\end{equation}
The coefficients $A$ and $N$ are shown in Table \ref{cspl}.

\begin{table}
\caption{Values of $N$ and $A$ for power law energy 
dependence of cross section high energy tail, Eq.(\ref{pl}), 
for cross sections presented in Fig. \ref{sigmaeg} }
  \label{cspl}
\begin{tabular}{|c|c|c| } 
\hline
Reaction&$N$ &$A$ (mb)   \\
  \hline
$\gamma e\leftrightarrow \gamma e$&- 1.7&330 \\
$\gamma\gamma \leftrightarrow e^+e^-$& -1.7 &410 \\
$ e^+e^- \leftrightarrow \gamma\gamma$ & -1.7 &309 \\
\hline
\end{tabular}
\end{table}

The decrease of cross sections with particles energy
increase could result in the early escape (freeze-out) of very high energy particles from
the plasma which would  impact   heavier particle (muon, meson) production. However, this
has no material impact on our present considerations.

\subsection{Lorentz-invariant pair production and annihilation rates}\label{plpr2C}

The equation for pair production and annihilation rate 
$R_{\gamma\gamma \leftrightarrow ee}$ ($R_{12\leftrightarrow 34}$) 
in Eq.(\ref{lgg}) is similar to the equation for the rate of muon 
production in photons fusion that we presented
in~\cite{Kuznetsova:2008jt}:
\begin{eqnarray}
&&R_{\gamma \gamma \leftrightarrow ee}=
\frac{g_{\gamma}^2}{2(2\pi)^8}\int\frac{d^{3}{p_{1}^{\gamma}}}{2E_{1}^{\gamma}} \int\frac{d^{3}{p_{2}^
{\gamma}}}{2E_{2}^{\gamma}}
\int\frac{d^{3}{p_{3}^{e}}}{2E_{3}^{e}}
\int\frac{d^{3}{p_{4}^{e}}}{2E_{4}^{e}}\times
  \nonumber\\[0.2cm]
&&\times \delta^{4}\left(p_{1}^{\gamma} + p_{2}^{\gamma} -
p_{3}^{e} - p_{4}^{e} \right) \sum_{\rm spin}\left|\langle
p_{1}^{\gamma}p_{2}^{\gamma}\left| M_{\gamma\gamma \rightarrow
ee}\right|p_{3}^{e}p_{4}^{e}\rangle\right|^{2} \nonumber\\[0.2cm]
&&\times f_{\gamma}(p_{1}^{\gamma})f_{\gamma}(p_{2}^{\gamma})f_{e}(p_{3}^{e})f_{4}(p_{4}^{e})
\Upsilon_{\gamma}^{-2}\Upsilon_{e}^{-2}e^{u \cdot(p_{1}^{\gamma} +
p_{2}^{\gamma} )/T},\label{ggee}
\end{eqnarray}

We define
\begin{eqnarray}
{\bf q} = p_1+p_2;\,\,\,\,{\bf p}=\frac{1}{2}(p_1-p_2); \nonumber\\
{\bf q}^{\prime}=p_4 + p_3;\,\,\,\,{\bf p^{\prime}}=\frac{1}{2}(p_4-p_3);\label{qpp}
\end{eqnarray}
and then we have ${\bf q}^2=q_0^2-q^2 = q_0^{\prime\,2}-q^{\prime\,2}=s
\geq 2m^2$ and
\begin{eqnarray}
p_1 = \frac{{\bf q}}{2}+{\bf p};\qquad p_2 = -{\bf p}+\frac{1}{2}{\bf q}; \nonumber\\
p_3 = \frac{{\bf q}}{2}-{\bf p^{\prime}};\qquad p_4 = \frac{{\bf q}}{2}+{\bf p}^{\prime}. \label{p1234}
\end{eqnarray}

Using
\begin{equation}
\int\frac{d^3p}{2E}=\int d^4p\delta(p^2-m^2)\theta(p_0)\equiv 
\int d^4p\delta_0(p^2-m^2),
\end{equation}
and $p^2_{3,4}-m^2=0$ and $p_{1, 2}^2=0$, we obtain:

\begin{eqnarray}
 {R_{ \gamma\gamma \leftrightarrow ee}}&=& \frac{g_{\gamma}^2}{2(2\pi)^8} \int
d^4q\int d^4p\int d^4p^{\prime}
\delta_0\left({p}_1^2\right)\delta_0\left({p}_2^2\right) 
\nonumber\\[0.3cm]
&&\hspace*{-1.2cm}\times
\delta_0\left({p}_3^2-m^2\right)
\delta_0\left({p}^2_4-m^2\right)
\Upsilon_{\gamma}^{-2}\Upsilon_{e}^{-2}\exp{(q_0/T)}
\nonumber\\[0.3cm]
&&\hspace*{-1.3cm}\times \sum|M_{  \gamma \gamma \rightarrow ee}|^2
  f_{\gamma}\left(p_1^0\right)f_{e}\left(p_3^0\right)
f_{\gamma}\left(p_2^0\right) f_{e} \left(p_4^0\right). 
\label{rategg}
\end{eqnarray}

The integrals from Eq.(\ref{rategg}) can be evaluated in spherical
coordinates. The angle coordinates are chosen with respect to the
direction of 
$\overrightarrow{q}=\overrightarrow{p_3}+\overrightarrow {p_1}$:
$$q_{\mu}=(q_0,0,0,q),\,\,\,\,p_{\mu}=(p_0, p\sin\theta,0, p\cos\theta),$$
$$p_{\mu}^{\prime}=(p^{\prime}_0, p^{\prime}\sin\phi\sin\chi, p^{\prime}
\sin\phi\cos\chi, p^{\prime}\cos\phi).$$

Using Eqs. (\ref{p1234}) and delta functions we obtain the equations:
\begin{eqnarray}
&&p_0^2-p^2 +\frac{s}{4} + p_0q_0 - pq\cos{(\theta)}-m^2= 0; \label{pg1}\\
&&p_0^2-p^2+\frac{s}{4} - p_0q_0 + pq\cos{(\theta)}-m^2= 0; \label{pg2}\\
&&p^{\prime\,2}_0-p^{\prime\,2} + \frac{s}{4} + p^{\prime}_0q_0 - p^{\prime}q\cos{(\phi)} = 0; \label{pprg1}\\
&&p^{\prime\,2}_0-p^{\prime\,2}+\frac{s}{4}-p^{\prime}_0q_0 + p^{\prime}q\cos{(\phi)} = 0; \label{pprg2}
\end{eqnarray}
then using  $\delta(f(x))=\sum_i 1/|f^{\prime}(x_i)|\delta(x-x_i)$, we can rewrite the integral (similar to~\cite{Matsui:1985eu}) as
\begin{widetext}
\begin{eqnarray}
 {R_{\gamma\gamma \leftrightarrow ee}}&=&
\frac{g_{\gamma}^2}{(2\pi)^6 16} \int_{2m}^{\infty}dq_0
\int_0^{\sqrt{q_0^2-s}}dq\int_{q_1}^{q_2}dp_0\int_{q^*_1}^{q^*_2}dp^{\prime}_0
\int_0^{\infty}dp\int_0^{\infty}dp\prime\int^{1}_{-1}d(\cos{\theta})\int^{1}_{-1}d(\cos{\phi})\
\int_0^{2\pi}d{\chi} \sum|M_{\gamma\gamma \rightarrow ee}|^2\nonumber\\[0.4cm]
&&\times
\delta\left(p-\left(p_0^2+\frac{s}{4}-{m^2}\right)^{1/2} \right)
\delta\left(p^{\prime}-\left(p^{\prime\,2}_0+\frac{s}{4}\right)^{1/2}\right)
\delta\left(\cos{\phi}-\frac{q_0p^{\prime}_0}{qp^{\prime}}\right) 
\delta\left(\cos{\theta}-\frac{q_0p_0}{pq}\right)\nonumber\\[0.4cm]
&&\times f_{\gamma}\left(\frac{q_0}{2}+p_0^{\prime}\right)f_{\gamma}\left(\frac{q_0}{2}-p_0^{\prime}\right)\Upsilon_{e}^{-2}
\Upsilon_{\gamma}^{-2}f_{e}\left(\frac{q_0}{2}+p_0\right)f_{e}
\left(\frac{q_0}{2}-p_0\right)\exp{(q_0/T)}, \label{rateee}
\end{eqnarray}
\end{widetext}
where
\begin{eqnarray}
&&q_{1,2}=\pm \frac{q}{2}\sqrt{1-\frac{4m^2}{s}};\\
&&q^*_{1,2}=\pm \frac{q}{2}.
\end{eqnarray}
$q_{1,2}$ and $q^*_{1,2}$ come from constrains $\cos{\theta}$, $\cos{\phi}<1$ and $\delta$ functions in Eq.(\ref{rateee}).

The integration over $p$, $p^{\prime}$, $\cos{\theta}$, and $\cos{\phi}$
can be done analytically considering $\delta$ functions. Other
integrals can be evaluated numerically. In order to simplify numerical integration we introduce dimensionless
variables:
\begin{eqnarray}
q = (q_0^2-4m^2)^{1/2}z; \label{z0}\\
\nonumber\\
p_0= \frac{q}{2}\sqrt{1-\frac{4m^2}{s}}x; \label{x0}\\
\nonumber\\
p_0^{\prime}=\frac{q}{2}y; \label{y0},
\end{eqnarray}
$0<z<1$; $-1<x(y)<1$.

In these variables, using the $\delta$ function in Eq.(\ref{rateee}), we obtain for $u$ and $t$:
\begin{eqnarray}\label{ut1}
&&u = (p-p^{\prime})^2={m^2}-\frac{s}{2}+\frac{s}{2}\sqrt{1-\frac{4m^2}{s}}\times\nonumber\\
&&(xy - \sqrt{(1-x^2)(1-y^2)}\sin{\chi});\nonumber\\
\nonumber\\
&&t = (p+p^{\prime})^2=m^2-\frac{s}{2}-\frac{s}{2}\sqrt{1-\frac{4m^2}{s}}\times\nonumber\\
&&(xy-\sqrt{(1-x^2)(1-y^2)}\sin{\chi}).\\
\nonumber
\end{eqnarray}

Using new variables (\ref{z0})-(\ref{y0}) the Eq. (\ref{ggee}) is
\begin{eqnarray}
R_{\gamma\gamma \rightarrow ee} &=& \frac{g_{\gamma}^2}{2^{12}\pi^6}\int_{2m}^{\infty}dq_0
\exp(q_0)\int_0^1dzz^2(q_0^2-4m^2)^{3/2} \nonumber\\[0.3cm]
&&\hspace*{-0.9cm}\times\int_0^{2\pi} d\chi\int_{-1}^1 dx\int_{-1}^1dy\sqrt{1-\frac{4m^2}{s}}\sum|M_{\gamma\gamma \rightarrow ee}|^2
 \nonumber\\[0.3cm]
&&\hspace*{-0.4cm} \times  \Upsilon_{e}^{-2}f_{\gamma}
\left(\frac{q_0}{2}+p^{\prime}_0\right)f_{\gamma}\left(\frac{q_0}{2}-p^{\prime}_0\right) \nonumber\\[0.3cm]
&&\hspace*{-0.4cm} 
\times\Upsilon_{\gamma}^{-2}f_{e}\left(\frac{q_0}{2}+p_0\right)f_{e}
\left(\frac{q_0}{2}-p_0\right).\label{rateggf}
\end{eqnarray}
The matrix element is defined by Eq.~\ref{m2ggee}.

\subsection{Compton scattering Lorentz-invariant rate}  \label{plpr2D}

The equation for the Compton scattering rate is
\begin{eqnarray}
&&R_{\gamma e}=
\frac{g_eg_{\gamma}}{(2\pi)^8}\int\frac{d^{3}{p_{3}^{e}}}{2E_{3}^{e}} 
\int\frac{d^{3}{p_{1}^ {\gamma}}}{2E_{1}^{\gamma}}
\int\frac{d^{3}{p_{3}^{e}}}{2E_{4}^{e}}
\int\frac{d^{3}{p_{2}^{\gamma}}}{2E_{2}^{\gamma}}
 \label{comscat} \\[0.2cm]
&&\times\delta^{4}\left(p_{1}^{\gamma} + p_{3}^{e} -
p_{4}^{e} - p_{2}^{\gamma} \right) \sum_{\rm spin}\left|\langle
p_{1}^{\gamma}p_{3}^{e}\left| M_{\gamma e \rightarrow \gamma
e}\right|p_{2}^{\gamma}p_{4}^{\,e}\rangle\right|^{2}\nonumber\\
&&\times f_{\gamma}(p_{1\gamma})f_{\gamma}(p_{2\gamma})f_{e}(p_{3}^{e})f_{4}(p_{4}^{e})
\Upsilon_{\gamma}^{-2}\Upsilon_{e}^{-2}e^{u \cdot (p_{3}^{e} +
p_{1}^{\gamma} )/T}\nonumber 
\end{eqnarray}

The Feynman diagrams for Compton scattering are shown in figure~\ref{eeprod} (b).
The matrix element for Compton scattering, Eq.(\ref{m2comp}), can be obtained by crossing $t$ and $s$, Eq.(\ref{stcr}),
in the matrix element for pair production and annihilation, Eq.(\ref{m2ggee}). 

In this case $s = ({\bf p_3}+{\bf p_1})^2$, $t=({\bf p_1}-{\bf p_2})^2$ and $u=({\bf p_3}-{\bf p_2})^2$.

We need to cross $p_2$ and $p_3$ in variables $q$, $p$ and $p^{\prime}$ definition, Eq.(\ref{qpp}). 

In this case we have ${\bf q}^2=q_0^2-q^2 = q_0^{\prime\,2}-q^{\prime\,2}=s
\geq m^2$ and
\begin{eqnarray}
p_1 = \frac{{\bf q}}{2}+{\bf p};\qquad p_2 = -{\bf p}^{\prime}+\frac{1}{2}{\bf q}; \nonumber\\
p_3 = \frac{{\bf q}}{2}-{\bf p};\qquad p_4 = \frac{{\bf q}}{2}+{\bf p}^{\prime}. \label{p1324}
\end{eqnarray}

Then, similar to pair production and annihilation rate, we obtain:
\begin{widetext}
\begin{eqnarray}
 {R_{e \gamma }}&=& \frac{g_eg_{\gamma}}{(2\pi)^8} \int
d^4q\int d^4p\int d^4p^{\prime}
\delta\left({p}_1^2\right)\delta\left({ p}_3^2-m^2\right)\delta\left({p}^2_4-m^2\right)\delta\left({
p}_2^2\right)\theta\left({p}^0_1\right) \theta\left(
p_2^0\right)\theta\left({p}_3^0\right)
\theta\left({ p}_4^0\right)\nonumber\\[0.3cm]
&&\times \sum|M_{\gamma e \rightarrow \gamma
e}|^2\Upsilon_{e}^{-2}f_{e}\left(p_3^0\right)f_{\gamma}\left(p_1^0\right)
\Upsilon_{\gamma}^{-2}f_{\gamma}\left(p_2^0\right) f_{e} \left(p_4^0\right)\exp{(q_0/T)}. \label{ratege}
\end{eqnarray}
\end{widetext}

Using Eqs. (\ref{p1324}) and $\delta$ functions from Eq.(\ref{ratege}) we obtain the equations:
\begin{equation}
p_0^2-p^2 +\frac{s}{4} + p_0q_0 - pq\cos{(\theta)}= 0; \label{p1}
\end{equation}
\begin{equation}
p_0^2-p^2+\frac{s}{4} - p_0q_0 + pq\cos{(\theta)}-m^2= 0; \label{p2}
\end{equation}
\begin{equation}
p^{\prime\,2}_0-p^{\prime\,2} + \frac{s}{4} + p^{\prime}_0q_0 - p^{\prime}q\cos{(\phi)}-m^2 = 0; \label{ppr1}
\end{equation}
\begin{equation}
p^{\prime\,2}_0-p^{\prime\,2}+\frac{s}{4}-p^{\prime}_0q_0 + p^{\prime}q\cos{(\phi)} = 0; \label{ppr2}
\end{equation}
then using  $\delta(f(x))=\sum_i 1/|f^{\prime}(x_i)|\delta(x-x_i)$, we can rewrite the integral as
\begin{widetext}
\begin{eqnarray}
&& {R_{e\gamma}}=
\frac{2g_eg_{\gamma}}{(2\pi)^6 16} \int_{m}^{\infty}dq_0
\int_0^{\sqrt{q_0^2-s}}dq\int_{q_1}^{q_2}dp_0\int_{q^*_1}^{q^*_2}dp^{\prime}_0
\int_0^{\infty}dp\int_0^{\infty}dp\prime\int^{1}_{-1}d(\cos{\theta})\int^{1}_{-1}d(\cos{\phi})\
\int_0^{2\pi}d{\chi} \nonumber\\
&&\times\sum|M_{e
\gamma \rightarrow e\gamma}|^2\delta\left(p-\left(p_0^2+\frac{s}{4}-\frac{m^2}{2}\right)^{1/2} \right)
\delta
\left(p^{\prime}-\left(p^{\prime\,2}_0-\frac{m^2}{2}+\frac{s}{4}\right)^{1/2}\right)\delta\left(\cos{\phi}-\frac{q_0p^{\prime}_0}{qp^{\prime}}+\frac{m^2}{2qp^{\prime}}\right) \nonumber\\[0.4cm]
&&\times \delta\left(\cos{\theta}-\frac{q_0p_0}{pq}-\frac{m^2}{2qp}\right)f_{e}\left(\frac{q_0}{2}-p_0\right)f_{\gamma}\left(\frac{q_0}{2}+p_0
\right)\Upsilon_{e}^{-2}\Upsilon_{\gamma}^{-2}f_{\gamma}\left(\frac{q_0}{2}-p^{\prime}_0\right)f_{e}
\left(\frac{q_0}{2}+p^{\prime}_0\right)\exp{(q_0/T)}, \label{ratecsc}
\end{eqnarray}
\end{widetext}
where in this case 
\begin{eqnarray}
&&q_{1,2}=-\frac{m^2q_0}{2s}\pm \frac{q}{2}{\left(1-\frac{m^2}{s}\right)};\\
&&q^*_{1,2}=\frac{m^2q_0}{2s}\pm \frac{q}{2}{\left(1-\frac{m^2}{s}\right)}.
\end{eqnarray}

The integration over $p$, $p^{\prime}$, $\cos{\theta}$ and $\cos{\phi}$
can be done analytically considering $\delta$ functions. Other
integrals can be evaluated numerically. In order to simplify numerical integration we introduce dimensionless
variables:
\begin{eqnarray}
q = (q_0^2-m^2)^{1/2}z; \label{z}\\
\nonumber\\
p_0= -\frac{m^2q_0}{2s} + \frac{q}{2}\left(1-\frac{m^2}{s}\right)x, \label{x}\\
\nonumber\\
p_0^{\prime}=\frac{m^2q_0}{2s} + \frac{q}{2}\left(1-\frac{m^2}{s}\right)y \label{y},
\end{eqnarray}
$0<z<1$; $-1<x(y)<1$.

With these variables, using the $\delta$ function in Eq.(\ref{ratecsc}), we obtain for $u$ and $t$:
\begin{eqnarray}\label{ut2}
 u & =& (p-p^{\prime})^2={m^2}-\frac{s}{2}+\frac{m^4}{2s}+\frac{s}{2}\left(1-\frac{m^2}{s}\right)^2 \nonumber\\
&&\times(xy - \sqrt{(1-x^2)(1-y^2)}\sin{\chi});\nonumber\\
\nonumber\\
 t & =& (p+p^{\prime})^2={m^2}-\frac{s}{2}-\frac{m^4}{2s}-\frac{s}{2}\left(1-\frac{m^2}{s}\right)^2 \nonumber\\
&&\times(xy-\sqrt{(1-x^2)(1-y^2)}\sin{\chi}).
\end{eqnarray}
Then limits for $t$ are $0 > t >-s+2m^2-m^4/s$.
In the limit $s>>m^2$ the variables $u$ and $t$ for Compton scattering go to the same limit as $u$ and $t$ for pair production and annihilation,
Eq.(\ref{ut1}).

Using new variables (\ref{z})-(\ref{y}) the Eq. (\ref{comscat}) is
\begin{widetext}
\begin{eqnarray}
 R_{e\gamma} &=& \frac{g_{\gamma}g_e}{2^{11}\pi^6}\int_{m}^{\infty}dq_0
\exp(q_0)\int_0^1dzz^2(q_0^2-m^2)^{3/2}\int_0^{2\pi} d\chi
\int_{-1}^1 dx\int_{-1}^1dy\left(1-\frac{m^2}{s}\right)^2\sum|M_{\gamma e^{\pm}\rightarrow \gamma e^{\pm}}|^2
 \nonumber\\[0.4cm]
&& \times  \Upsilon_{e}^{-2}f_{e^{\pm}}
\left(\frac{q_0}{2}-p_0\right)f_{\gamma}\left(\frac{q_0}{2}+p_0\right)\Upsilon_{\gamma}^{-2}f_{\gamma}\left(\frac{q_0}{2}-p^{\prime}_0\right)f_{e}
\left(\frac{q_0}{2}+p^{\prime}_0\right).\label{ratecomf}
\end{eqnarray}
\end{widetext}

\subsection{M{\o}ller and Bhabha scatterings}\label{plpr2E}

In this section we consider M{\o }ller and Bhabha scatterings, Eq.(\ref{Moller}) and Eq.(\ref{bhaba}), respectively.
To remove Coulomb divergence in reaction rates we introduce thermal photon mass induced by Debye screening~\cite{Thoma:2008my}:
\begin{equation}
m_{\gamma} = \omega_{pl},
\end{equation}
where $\omega_{pl}$ is plasma frequency.
\begin{equation}
m_{\gamma}^2=\frac{4e^2}{3\pi^2}\int_0^{\infty}{dp p f_F(p)}.
\end{equation} 
For highly relativistic plasma $T>>m_e$
\begin{equation}
m_{\gamma} = \frac{\sqrt{4\pi\alpha}}{3}T.
\end{equation}
The corresponding thermal mass of electron for high temperatures is~\cite{Thoma:2008my}
\begin{equation}
m = \frac{\sqrt{4\pi\alpha}}{2\sqrt{2}}T.
\end{equation} 
In our range of temperatures we assume:
\begin{equation}
m^2 =m_e^2 + \frac{4\pi\alpha}{8}T^2.
\end{equation}
In this section $m_e=0.5$ MeV is electron mass without thermal effects. 

The Feynman diagrams for M{\o}ller (a) and Bhabha scatterings (b) are shown in Fig.~\ref{eediagramsp} and the matrix elements are defined by Eqs.(\ref{M_fi_m}) and Eq.(\ref{M_fi_b}), respectively. 
In both cases all initial and final particles are massive and have the same mass. After transformations similar to the ones used for Compton scattering
the equation for the M{\o}ller (Bhabha) rate is
\begin{widetext}
\begin{eqnarray}
{R_{ee}} & = &
\frac{1}{1+I}\frac{2g_e^2}{(2\pi)^6 16} \int_{2m}^{\infty}dq_0
\int_0^{\sqrt{q_0^2-s}}dq\int_{q_1}^{q_2}dp_0\int_{q^*_1}^{q^*_2}dp^{\prime}_0
\int_0^{\infty}dp\int_0^{\infty}dp\prime\int^{1}_{-1}d(\cos{\theta})\int^{1}_{-1}d(\cos{\phi})\
\int_0^{2\pi}d{\chi} \nonumber\\[0.4cm]
&&\times\sum|M_{ee}|^2\delta\left(p-\left(p_0^2+\frac{s}{4}-{m^2}\right)^{1/2} \right)
\delta
\left(p^{\prime}-\left(p^{\prime\,2}_0-{m^2}+\frac{s}{4}\right)^{1/2}\right)\delta\left(\cos{\phi}-\frac{q_0p^{\prime}_0}{qp^{\prime}}\right) \nonumber\\[0.4cm]
&&\times \delta\left(\cos{\theta}-\frac{q_0p_0}{pq}\right)f_{e}\left(\frac{q_0}{2}+p_0\right)f_{e}\left(\frac{q_0}{2}-p_0\right)
\Upsilon_{e}^{-4}f_{e}\left(\frac{q_0}{2}+p^{\prime}_0\right)f_{e}
\left(\frac{q_0}{2}-p^{\prime}_0\right)\exp{(q_0/T)}, \label{ratemu1}
\end{eqnarray}
\end{widetext}
\begin{eqnarray}
&&q_{1,2}=\mp \frac{q}{2}\sqrt{\left(1-\frac{4m^2}{s}\right)};\\[0.3cm]
&&q^*_{1,2}=\mp \frac{q}{2}\sqrt{\left(1-\frac{4m^2}{s}\right)}.
\end{eqnarray}
$q_{1,2}$ and $q^*_{1,2}$ come from constraints $\cos{\theta}$, $\cos{\phi}<1$.
Then we introduce dimensionless 
variables similar to Eqs.(\ref{z})-(\ref{y}):
\begin{eqnarray}
q = (q_0^2-4m^2)^{1/2}z; \label{z1}\\[0.3cm]
p_0= \frac{q}{2}\sqrt{\left(1-\frac{4m^2}{s}\right)}x, \label{x1}\\[0.3cm]
p_0^{\prime}= \frac{q}{2}\sqrt{\left(1-\frac{4m^2}{s}\right)}y \label{y1},
\end{eqnarray}
$0<z<1$; $-1<x(y)<1$.

With these variables we obtain for $u$ and $t$:
\begin{eqnarray}\label{ut3}
u & = & (p-p^{\prime})^2=2m^2-\frac{s}{2}+\frac{s}{2}\left(1-\frac{4m^2}{s}\right) \nonumber\\
&&\times(xy - \sqrt{(1-x^2)(1-y^2)}\sin{\chi});\nonumber\\
\nonumber\\
 t & = & (p+p^{\prime})^2=2m^2-\frac{s}{2}-\frac{s}{2}\left(1-\frac{4m^2}{s}\right) \nonumber\\
&&\times(xy - \sqrt{(1-x^2)(1-y^2)}\sin{\chi}).
\end{eqnarray}

In new variables (\ref{z1})-(\ref{y1}) the equation for scattering rate (\ref{ratemu1}) is
\begin{widetext}
\begin{eqnarray}
 R_{ee} & =& \frac{1}{I+1}\frac{g_e^2}{2^{11}\pi^6}\int_{2m}^{\infty}dq_0
\exp(q_0)\int_0^1dzz^2(q_0^2-4m^2)^{3/2} 
\int_0^{2\pi} d\chi
\int_{-1}^1 dx\int_{-1}^1dy\left(1-\frac{4m^2}{s}\right)f_{e}
\left(\frac{q_0}{2}+p_0\right)\Upsilon_{e}^{-4} 
 \nonumber\\[0.5cm]
&&  \times\sum|M_{ee}|^2f_{e}\left(\frac{q_0}{2}-p_0\right)f_{e}\left(\frac{q_0}{2}+p^{\prime}_0\right)
f_{e}\left(\frac{q_0}{2}-p^{\prime}_0\right).\label{rateeef}
\end{eqnarray}
\end{widetext}

\section{Numerical Results and Discussion} \label{numres}
\subsection{Thermal cross section and Thompson limit} \label{Thompson}

\begin{figure}[h]
\centering
\includegraphics[width=8cm,height=8cm]{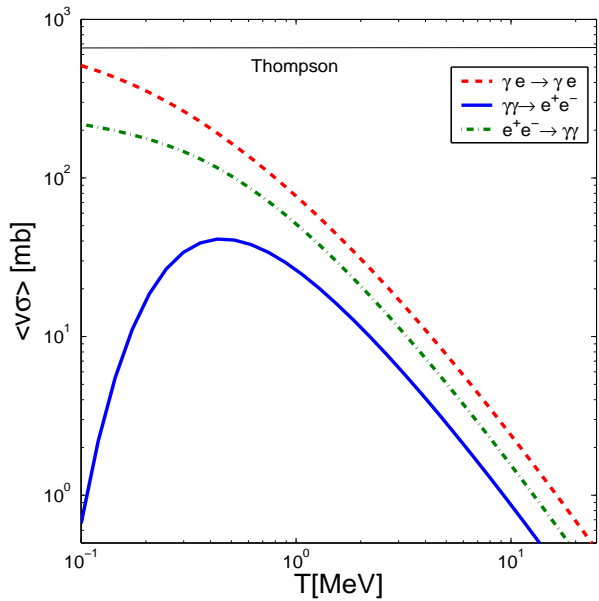}
\caption{\small{The thermally averaged cross sections 
$\left\langle v\sigma\right\rangle/c$ for Compton 
scattering (dashed red line), pair production  
(solid blue  line) and annihilation ( dash-dotted green line) 
in the observer rest frame shown as
a function of temperature $T$ at $\Upsilon=1$}} \label{sigcom}
\end{figure}

The thermally averaged cross sections $\left\langle v\sigma\right\rangle/c$ 
in observer frames for pair production, annihilation and Compton scattering
 Eq.(\ref{ggee1}), (\ref{ge}) are evaluated using the Lorentz-invariant 
rates, Eq.(\ref{thcr}) with rates, evaluated numerically using 
Eq.(\ref{rateggf}) for pair production and annihilation 
and Eq.(\ref{ratecomf}) for Compton scattering. 
In Fig.~\ref{sigcom} we show $\left\langle v\sigma\right\rangle/c$ 
for Compton scattering (dashed red line), pair production (solid blue line) 
and pair annihilation (dash-dotted green line). For high $T$ we observe the power 
law fall-off of the thermal cross sections and the corresponding parameters
are presented in Table~\ref{csplt}. 
The power $N=1.73$ in power law for pair production/annihilation  and Compton scattering
is close for nonthermal cross section in this case. 
At high temperatures these cross sections  
are almost parallel. This is in agreement with $t$-averaged 
cross sections Eqs.(\ref{crcom})-(\ref{crggee}).

\begin{table}
\caption{Values of $N$ and $A$ for the power law temperature
dependence of thermal cross section high energy tail, see Eq.(\ref{pl}) for the shape assumed, 
for cross sections presented in figure  \ref{sigcom}.}
 \label{csplt}
\begin{tabular}{|c|c|c|} 
\hline
Reaction& $N$ & $A$ [mb]  \\
  \hline
$\gamma e\leftrightarrow \gamma e$ &-1.73& 120\\
$\gamma\gamma \leftrightarrow e^+e^-$ &-1.73&80\\
$ e^+e^- \leftrightarrow \gamma\gamma$ &-1.73&46\\
\hline
\end{tabular}
\end{table}

Considering electron production, the density of photons with energy 
larger than the threshold for pair production drops in the tail of 
Boltzmann distribution with decreasing temperature, therefore the 
thermal cross section $\left\langle v\sigma\right\rangle$ starts 
to decrease at $T < m$ and goes to 0 as $T \rightarrow 0$.  For
 electron-positron annihilation $\left\langle v\sigma\right\rangle/c$ 
stays for $T\rightarrow 0$, $v \rightarrow 0$, since $\sigma \propto 1/v$.

\begin{figure}[h]
\centering
\includegraphics[width=8cm,height=8cm]{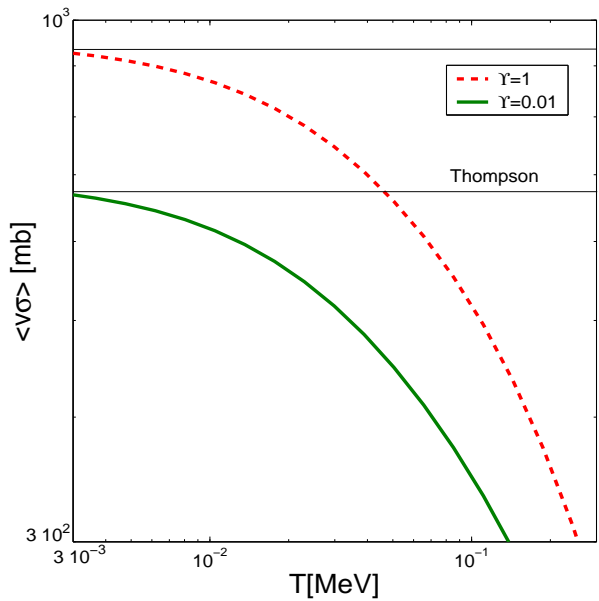}
\caption{\small{The thermally averaged cross sections 
$\left\langle v\sigma\right\rangle/c$ for Compton 
scattering for $\Upsilon_{\gamma}=1$ (dashed red line) and $\Upsilon_{\gamma} =0.01$
(solid green line) as a function of temperature $T$. 
}} \label{sigcomth}
\end{figure}

 At $T << m$, the Compton scattering cross section $\left\langle v\sigma\right\rangle$ 
is approaching the Thompson limit very slowly, which in case $\Upsilon=1$ is slightly 
[$\left\langle v\sigma\right\rangle_{comp}(\Upsilon=1)/c\sigma_{th}\approx 1.36$] 
above the actual classical Thompson cross section 
$$\sigma = \frac{8\pi \alpha^2}{3m^2}=6.7\,\times 10^2\,\rm mb.$$ 
This result is shown in Fig.~\ref{sigcomth} (dashed red line). 
This difference with the Thompson limit is due to quantum effects 
from massless photons. We can use Boltzmann distribution for photons 
only when $\Upsilon_{\gamma}<<1$. In this case we obtain that thermal Compton 
cross section goes to the Thompson limit (green solid line 
in Fig.~\ref{sigcomth}). In order to explain the difference 
between the cross section for $\Upsilon_{\gamma}=1$ and the Thompson limit, 
we assume that electrons are at rest at small $T$ and thermal 
photons are scattering without energy change $E_i=E_f$. 
Then the cross section quantum enhancement factor $Q$ 
is the thermal average from the Bose enhancement factor $f_{\gamma}(p)+1$:
\begin{equation}
Q=\frac{g_{\gamma}}{2\pi^2n_\gamma}\int_0^{\infty}f(p)(f(p)+1)p^2dp.
\end{equation}
In the Boltzmann limit $f(p)+1 \rightarrow 1$ and $Q \rightarrow 1$. 
For $\Upsilon_{\gamma} = 1$ we obtained $Q=1.38$.
Further discussion about the Boltzmann and the 
quantum limit comparison is found in Sec.~\ref{oldappendix}

\subsection{Mean Free Paths} \label{Paths}

\begin{figure*}
\centering
\includegraphics[width=8.6cm,height=8.5cm]{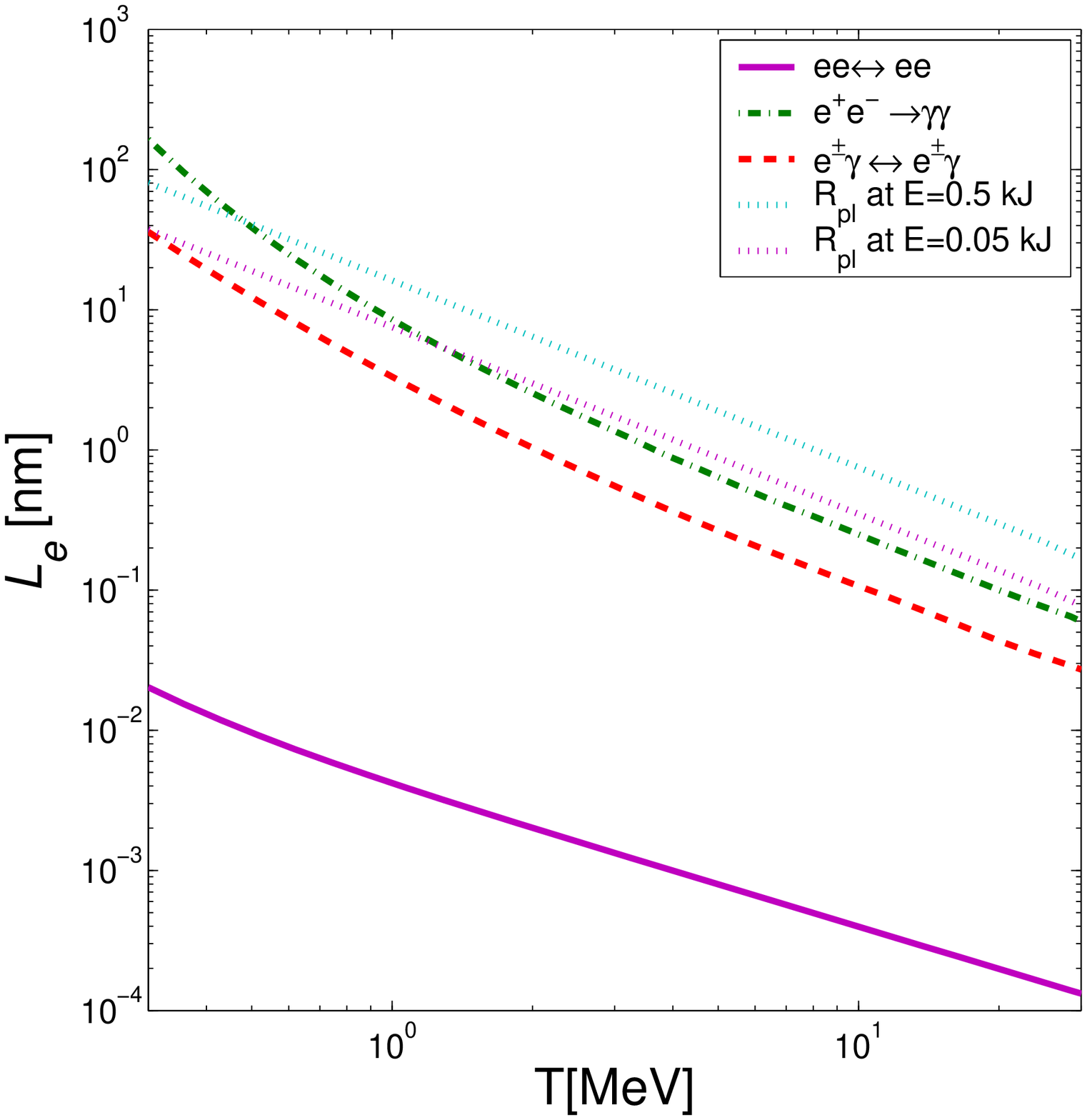}
\includegraphics[width=8.6cm,height=8.5cm]{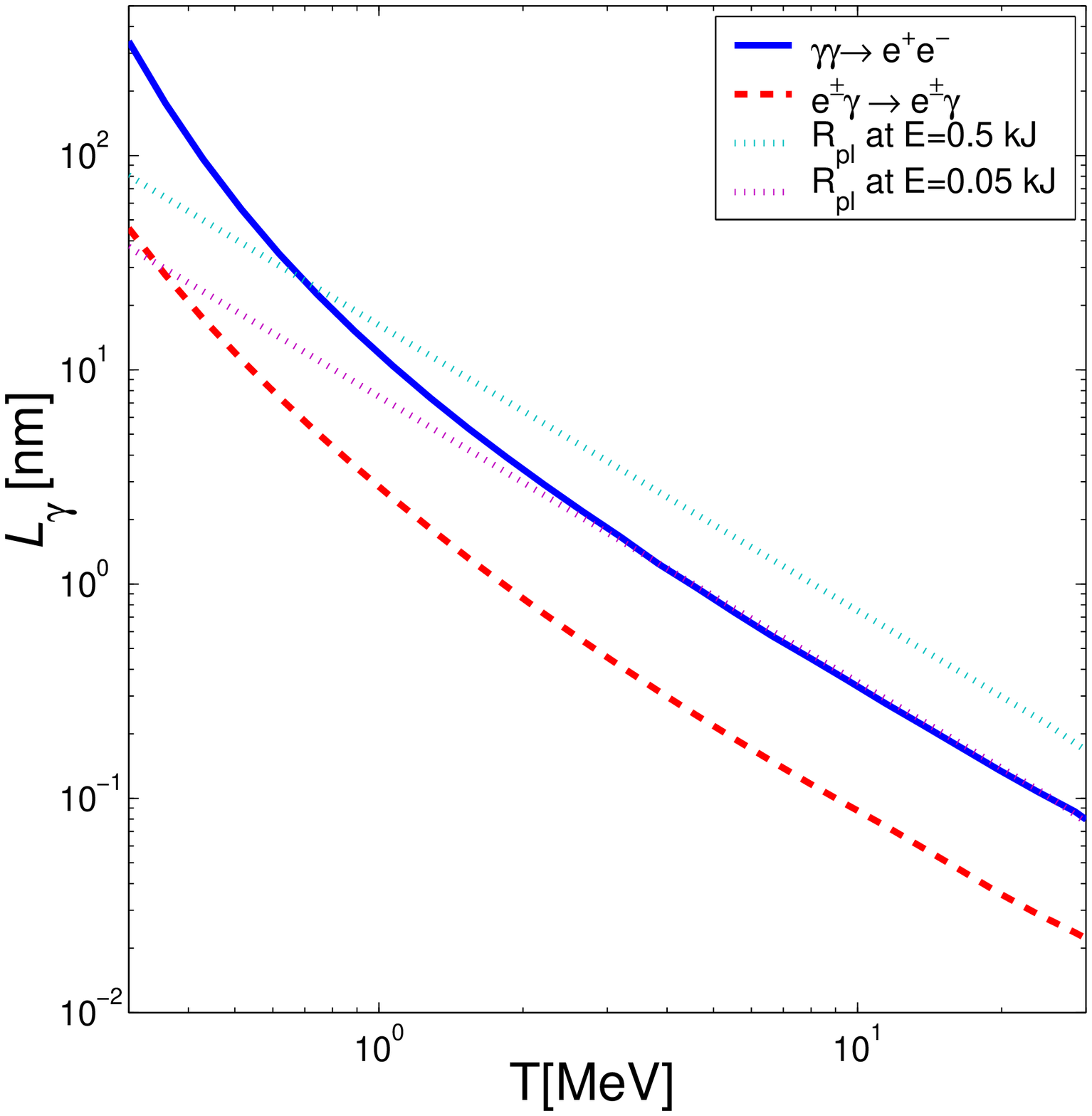}
\caption{\small{Left panel: the mean free paths of electron (positron) 
shown as functions of plasma temperature in reactions 
Eqs.(\ref{ggee1}), (\ref{ge})(red dashed, green dash-dot lines, respectively), 
and reactions Eqs.(\ref{Moller}) and (\ref{bhaba}) together 
(purple solid line); right panel: the mean free path of photon for Compton scattering and
pair production at $\Upsilon=1$ (red dashed and blue solid lines) as functions of
temperature $T$;  radius of equilibrium
($\Upsilon=1$) plasma at energy 500 J (upper, light blue, dotted line) 
and at energy 50 J (lower, purple, dotted line) as a function of $T$
on both sides.}} \label{tauee1}
\end{figure*}

In Fig. ~\ref{tauee1} (left panel) we show the mean free paths of
 electron (positron) in EP$^3$ plasma in pair annihilation and 
Compton scattering reactions Eqs.(\ref{ggee1}) and (\ref{ge})
(green dash-dotted and red dashed lines, respectively), and 
in M{\o}ller and Bhabha scattering reactions Eqs.(\ref{Moller}) and (\ref{bhaba}) 
together (purple solid line) ($\Upsilon_e = \Upsilon_{\gamma} =\Upsilon= 1$). 
The mean free paths are evaluated according to Eq.(\ref{lgg}): 
$$L_{ee \rightarrow \gamma\gamma} = \frac{n_e}{R_{{ee 
   \leftrightarrow \gamma\gamma}}};\qquad L_{e\gamma} 
    = \frac{n_e}{R_{e\gamma}};\qquad L_{ee
\rightarrow ee} = \frac{n_e}{R_{{ee}}}$$
with rates evaluated, using Eq.(\ref{rateggf}) for pair annihilation, 
Eq.(\ref{ratecomf}) for Compton scattering, and Eq.(\ref{rateeef}) 
for M{\o}ller and Bhabha scatterings. To determine the possible 
radius of opaque and equilibrated plasma [equivalent to opaqueness 
condition, Eq.~(\ref{opcon}), being satisfied] at a given energy, 
we show the radius of a plasma drop as a function of temperature 
[using Eq.(\ref{SB})] at expected experimental energies 500 J 
(upper light blue dotted line) and 50 J (lower purple dotted line).

The mean free path of electron or positron in M{\o}ller and 
Bhabha scattering remains much below the equilibrium plasma radius   
at energy 500 J and the temperature range considered here, 
up to the smallest temperature considered $T \cong 0.3$ MeV. 
Therefore electrons and positrons are thermally equilibrated 
at our temperature range. The $L_{e}$ in M{\o}ller and Bhabha 
scattering is much smaller than $L_{e}$ in Compton scattering
and in pair annihilation. Therefore electron and positron become 
thermally equilibrated first after their production by the laser field 
and then they produce photons over the larger time scale. 

We do not continue calculations for smaller temperatures since
 for ($T<<m_e$) it becomes impossible to create equilibrium 
plasma with zero chemical potential considered here, 
and hence for this reason our calculations of M{\o }ller and Bhabha scattering 
rates are developed for temperatures $T>>m$. 

In Fig.~\ref{tauee1} (right pannel) the mean free photon 
paths $L_{\gamma}$ are shown for $e^+e^-$
pair production reaction (solid blue line) 
and Compton scattering (dashed red line)
as functions of temperature $T$ for plasma 
in chemical equilibrium ($\Upsilon_i = 1$). 
Equation (\ref{lgg}) is used together with Eqs. (\ref{rateggf}) 
and (\ref{ratecomf}) for pair production and Compton scattering rates:
$$L_{\gamma\gamma \rightarrow ee} = 
\frac{n_{\gamma}}{R_{{ee \leftrightarrow \gamma\gamma}}}; \qquad L_{\gamma e} 
    = \frac{n_{\gamma}}{R_{e\gamma}}.$$

The mean free paths for photon and electron scattering presented 
in Fig.~\ref{tauee1} have power law distribution at high temperature, $T>>m$,
\begin{equation}
L=A\left(\frac{T}{1\,{\rm MeV}}\right)^{N}. \label{lpl}
\end{equation}
The coefficients $A$ and $N$ for mean free paths are listed in Table \ref{lgA}.

We see that $L_{\gamma e}$ is less than $L_{\gamma \gamma}$ at $\Upsilon = 1$ 
over the full range of temperatures considered.
The line for $L_{\gamma e}$ follows that for 
$L_{\gamma \gamma}$ at $T>m_e$ and $L_{\gamma e}$ 
is about 4 times smaller than $L_{\gamma \gamma}$,
when values of $L_{e\gamma }$ are about twice as 
low as $L_{ee\leftrightarrow \gamma\gamma}$ at high temperatures. 
The factor making  $L_{\gamma e}$ to be half the length $L_{e\gamma }$
 is the photons scattering on both electrons and positrons.

\begin{table}
\caption{The values of $N$ and $A$ in the power law temperature 
dependence of free mean path high energy tail, Eq.(\ref{lpl})}
  \label{lgA}
  
\begin{tabular}{|c|c|c|}
\hline
     
reaction&$N$&$A$ [nm]\\
  \hline
$L_{\gamma},\gamma e\leftrightarrow \gamma e$& - 1.3& 1.8 \\
$L_{\gamma},\gamma\gamma \leftrightarrow e^+e^-$& -1.3& 6.5\\
$L_{e},\gamma e\leftrightarrow \gamma e$& - 1.3 & 5 \\
$ L_{e}, e^+e^- \leftrightarrow \gamma\gamma$ &- 1.3 &2\\
$ L_{e}, ee \leftrightarrow ee$ &- 1 &0.004\\
\hline
\end{tabular}
\end{table}

At small temperatures, $T \leq m$, the photon mean free path $L_{\gamma}$ 
drops with temperature faster than the plasma radius at a fixed energy. 
Plasma loses opaqueness when $L_{\gamma e}$ approaches the plasma radius in magnitude.
For $T \approx 0.5$ MeV the free photon path $L_{\gamma e}$ is much smaller 
than radius of plasma, $R=35$ nm, at energy $500$ J. Since plasma is opaque,
photons will stay in plasma volume for a long time.  For $T < 0.5$ MeV plasma 
becomes nonopaque and can lose energy, radiating photons. The plasma, which is 
opaque for photons, cannot have a radius larger than $R \approx 35$ nm for the 
considered energy content. In Table~\ref{ETR} we show the values of maximum 
radius, minimum temperature and corresponding $e^+e^-$ pair and photon 
densities for opaque plasma at fixed energies 500 and 50 J.

Because at high temperatures the mean free paths decrease slower than the 
plasma radius, $R \propto T^{4/3}$. Therefore there is a limit on the minimum energy need to
create opaque plasma. A similar energy limit also exists for chemically equilibrated plasma,
condition Eq.(\ref{cheqcon1}). The energy $E=500 J$ is not large enough 
to satisfy   this equilibration condition exactly, but it is close to the minimum 
of required energy.  The largest ratio $R/L_{\gamma\gamma} \approx 2$ arises
at temperatures $T \approx 4-6$ MeV (Fig. \ref{tauee1}, right panel) and plasma 
may be close to chemical equilibrium at this temperature range. 
The corresponding radius is $\approx 2$ nm.  The pair production or 
annihilation relaxation time Eq. (\ref{taugg}) is approximately 
$10^{-3}$ fs at $T=5$ MeV and the corresponding plasma lifespan 
must be larger than that to satisfy the chemical equilibration condition. 

\begin{table}
\caption{The values of maximum radius, minimum temperature and 
corresponding $e^+e^-$ pair and photon densities for opaque 
plasma at fixed energies 500 and 50 J}
 \label{ETR}
  
\begin{tabular}{|c|c|c|c|c|}
\hline
     & opaque& opaque &opaque &opaque  \\
   E [J]&$R_{max}$ [nm]&$T$ [MeV]&$n_{\gamma}$ [cm$^{-3}$]&$n_{e}$ [cm$^{-3}$]\\
  \hline
  &&&&\\
500 & 35 & 0.5 &$3.9 \times 10^{30}$&$5 \times 10^{30}$\\
50 & 3&2&$2.55 \times 10^{32}$&$3.8 \times 10^{32}$\\
\hline
\end{tabular}
\end{table}

In this work we considered the highest density of chemical 
equilibrium plasma at $\Upsilon_{e} =\Upsilon_{\gamma}=\Upsilon = 1$ 
(for photon opaqueness condition $\Upsilon_e=1$ is enough), 
because  the opaqueness condition Eq.(\ref{opcon}) is
quickly violated with $\Upsilon_i$ decreases. This is because 
the photon free path  $L_{\gamma} \propto 1/\Upsilon$, when the radius
of plasma $R_{\rm pl} \propto 1/\Upsilon^{1/3}$, Eq.(\ref{energ}). 
The result is $R_{\rm pl}/L_{\gamma} \propto \Upsilon^{2/3}$ 
at a fixed energy content.  Only for the relatively high plasma 
temperature is it possible to have a below equilibrium density plasma with  
small $\Upsilon$, without violation of opaqueness condition Eq.(\ref{opcon}).  
For example at $T=20$ MeV  and energy 500 J $R_{\rm pl}/L_{\gamma e}\approx 8$. 
In this case minimum possible 
$\Upsilon \approx 0.3$ and radius at fixed plasma energy increases
 by factor of $\approx 1.4$ only ($\approx$ 0.4 nm), 
as the result $R_{\rm pl}/L_{\gamma e}\approx 3$.

\subsection{Lorentz-invariant rates and Boltzmann limit}\label{oldappendix}

In order to evaluate the reaction rate in two - body processes 
in the relativistic Boltzmann (classical) limit, we can use reaction cross
section $\sigma(s)$, and the relation~\cite{Letessier:2002gp}:
\begin{equation}
{R_{1\,2 \rightarrow 34}} = \frac{g_1g_2}{32\pi^4}\frac{T}{1+I}
\int_{s_{th}}^{\infty}ds\sigma(s)\frac{\lambda_2(s)}{\sqrt{s}}K_1(\sqrt{s}/T),\label{ratebol}
\end{equation}
(however, compared to Ref.~\cite{Letessier:2002gp} Eq.(17.16) 
$R_{12\rightarrow 34} \rightarrow 
R_{12 \rightarrow 34}/(\Upsilon_1 \Upsilon_2)$),
where
\begin{equation}
\lambda_2(s) = (s-(m_{1}+m_2)^2)(s-(m_1-m_2)^2),
\end{equation}
$m_1$ and $m_2$, $g_1$ and $g_2$, $\Upsilon_1$ and $\Upsilon_2$ are masses, degeneracy and fugacities of initial interacting particles.

\begin{figure}[t]
\centering
\includegraphics[width=8.6cm,height=8.5cm, height=8.5cm]{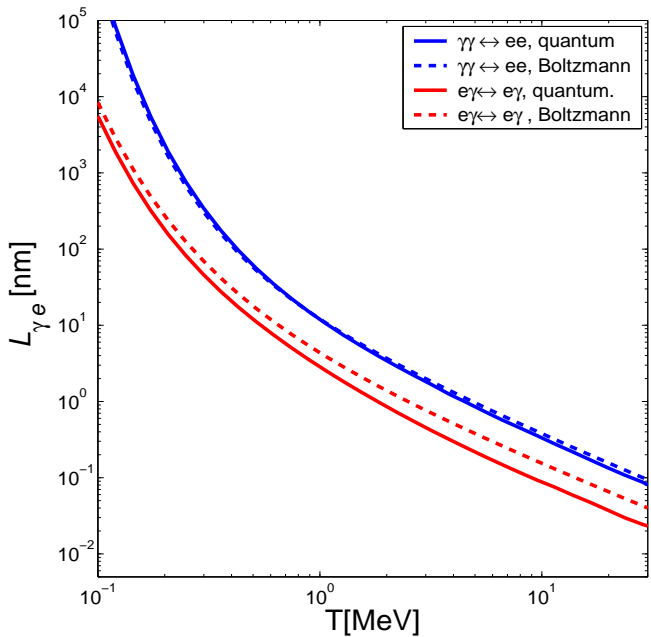}
\caption{\small{The photon mean free path for Compton scattering 
(lower red lines) and pair production (upper blue lines) 
obtained using reaction rate Eq.(\ref{ratecomf}) (solid lines)
 and using rate from Eq.(\ref{ratebol})(dashed lines), 
Boltzmann limit}} \label{taucomcomp}
\end{figure}

The cross sections can be evaluated using Eq.~(\ref{crcom})-(\ref{creegg}). In cases considered here 
Eq.~(\ref{ratebol}) with t-averaged cross section gives the same result for reaction rate 
as Eq.(\ref{m2ggee1}) with $\Upsilon \rightarrow 0$. In the general case the cross section
 averaging over t may also change the result compared 
to Eq.~(\ref{m2ggee1}). Calculating mean free path $L$ in the Boltzmann limit 
we also took the photon density in the Boltzmann limit as $n_{\gamma} \rightarrow n_{\gamma}/\Upsilon$ at
$\Upsilon \rightarrow 0$ to exclude all quantum distribution effects.

In Fig.~\ref{taucomcomp} we show our mean free paths for a photon 
subject to Compton or pair production process, 
for both quantum and Boltzmann (classical) limit 
as a functions of $T$. For Compton scattering quantum effects reduce the
mean free path at high density (i.e. high T). 
Somewhat smaller, but a noticeable quantum effect remains at  low temperature ($T < m$). 
For pair production there is no reduction of the mean free path at small temperature because of the threshold energy of photons in the pair production rate, and the smaller reduction for high T compared to Comptom scattering.

\section{Summary and Conclusion} \label{conc}

\begin{figure}
\centering
\includegraphics[width=8.6cm,height=8.5cm, height=8.5cm]{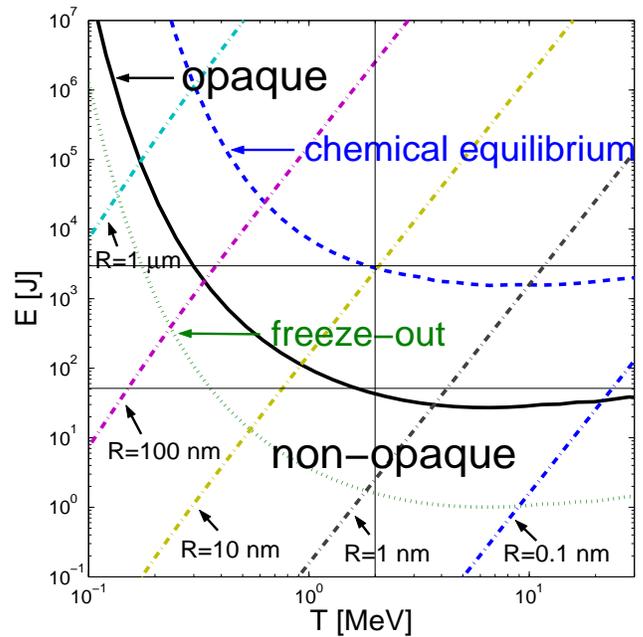}
\caption{\small{The energies of plasma drop with different radius 
($R=1\mu$m, 100 nm, 10 nm, 1 nm and 0.1 nm) as a functions of $T$ 
(dash-dotted lines - light blue, purple, brown, grey and blue, respectively); 
solid black line is energy vs. minimum $T$ of opaque plasma;
 dashed blue line is energy vs. minimum $T$ of chemically 
equilibrated plasma;)}} \label{energT}
\end{figure}

Current laser pulses can have energies up to 1 kJ and about 10 kJ
is expected to be produced in the foreseeable future~\cite{1kJ}.
Since we need to focus energy to sub diffraction size limit, $R<<\lambda$, only a part 
of the original energy is likely to be available to form plasma. 

The main purpose of this paper has been the evaluation of the domain of plasma 
size and temperature where thermally and chemically equilibrated 
EP$^3$ can exist at a given fireball energy. We evaluate the mean 
free path length $L$ for photons and electrons in EP$^3$ plasma, 
assuming thermal equilibrium and using the Lorentz invariant reaction rate.
This method allows us to take into account quantum effects in a dense medium.

Comparison of the electron (positron) mean free path with the plasma 
radius (figure~\ref{tauee1}, left) demonstrated that
electrons and positrons can be thermally equilibrated by M{\o}ller 
and Bhabha scattering at the temperature range that we considered. 
The small free path of electron and positron by M{\o}ller and Bhabha
scattering, compared to $e^+e^-$ pair annihilation, also means that  
electrons and positrons first become thermally equilibrated  and then 
they produce thermal photons. 

Comparison of photon mean free path $L_\gamma$ to plasma drop size  $R_{\rm pl}$
(see figure~\ref{tauee1}, right) indicates that we may have 
restrictions on maximum plasma size, minimum temperature and minimum energy of 
opaque and chemically equilibrated plasma, defined by opaqueness 
condition~Eq.(\ref{opcon}).
We find that the mean free path for $\gamma e$ Compton scattering 
is shorter than for pair production $\gamma \gamma\to e^+e^-$. Therefore  opaqueness
condition for photon will have $L_{\gamma}=L_{\gamma e\leftrightarrow\gamma e}$ 
and differs from chemical equilibration
condition Eq.(\ref{cheqcon1}).

The practical summary of our results is presented in figure~\ref{energT}. 
The energies of different size plasma drops are shown as a function of temperature.
The solid black line indicates the boundary below which the opaque plasma 
cannot exist; condition Eq.(\ref{opcon}) is violated. The blue dashed 
line indicates the boundary above which we have not only 
opaque but also chemically equilibrated plasma, condition Eq.(\ref{cheqcon1}). 
The green dotted line shows photon chemical freeze-out, $L_{\gamma}/R_{pl} = 1$.  
The straight lines show constrains given by plasma size between $E$ and $T$.

In Fig.~\ref{energT} we see that at fixed plasma size to have 
opaque plasma less energy is needed than to have chemically 
equilibrated plasma; for fixed energy the transition to 
chemically equilibrated plasma takes place at a smaller 
plasma sizes and higher temperatures and energies than the transition 
to opaque plasma. 
 
Only for $E>50$ J can we expect to form an opaque plasma for $T>2$ MeV. 
As more energy becomes available, the temperature limit drops and the 
radius increases ($R>5$ nm).  It is interesting to note that much more energy is needed 
to produce opaque plasma at lower compared to higher 
temperature, because plasma drop size has to increase 
considerably to satisfy the opaqueness condition Eq(\ref{opcon}).  Thus 
one of main objectives of laboratory experiments should be to seek paths to 
reduce the size of the domain in which energy deposition occurs. 

To summarize, 
our results indicate that chemically equilibrated, opaque, small -- size is about 5-50 nm -- plasma 
drop can be formed   for energy of a few  kJ which may be
experimentally attainable in foreseeable future.  Such a plasma drop would be an 
extraordinarily intense source of radiation and positrons. 

\vspace*{.2cm}
\subsubsection*{Acknowledgments}
This work was supported by the DFG Cluster of Excellence 
MAP (Munich Centre of Advanced Photonics), and by a grant from: the
U.S. Department of Energy  DE-FG02-04ER41318.

\vspace*{-0.3cm}


\end{document}